\documentclass[pre,twocolumn,superscriptaddress]{revtex4}  

\usepackage[paperwidth=210mm,paperheight=297mm,centering,hmargin=2cm,vmargin=2.5cm]{geometry}

\usepackage[pdftex]{graphicx}
\usepackage{amsmath,amsfonts,amssymb}
\usepackage{morefloats}
\usepackage{xcolor} 
\usepackage{tabularx}

\definecolor{bluemoi}{rgb}{0.25,0.50 ,0.75} 

\usepackage{hyperref}
\hypersetup{
  bookmarksopen=false,
  pdftitle=" ",
  pdfauthor=" ", 
  pdfsubject=" ", 
  pdftoolbar=true, 
  pdfmenubar=true, 
  pdfhighlight=/O, 
  colorlinks=true, 
  pdfpagemode=UseNone, 
  pdfpagelayout=SinglePage, 
  pdffitwindow=true, 
  linkcolor=bluemoi, 
  citecolor=bluemoi, 
  urlcolor=bluemoi 
}

\makeatletter
\renewcommand{\figurename}{Figure}
\renewcommand{\fnum@figure}{\small\textbf{\figurename~\thefigure}}
\renewcommand{\thefigure}{\arabic{figure}}
\makeatother

\makeatletter
\renewcommand{\tablename}{Table}
\renewcommand{\fnum@table}{\small\textbf{\tablename~\thetable}}
\renewcommand{\thetable}{\arabic{table}}
\makeatother

\begin{document}

\title{Deriving the number of jobs in proximity services from the number of inhabitants in French rural municipalities}

\author{Maxime Lenormand}\affiliation{IRSTEA, LISC, 24 avenue des Landais, 63172 AUBIERE, France}
\author{Sylvie Huet}\affiliation{IRSTEA, LISC, 24 avenue des Landais, 63172 AUBIERE, France}
\author{Guillaume Deffuant}\affiliation{IRSTEA, LISC, 24 avenue des Landais, 63172 AUBIERE, France}

\begin{abstract} 
We use a minimum requirement approach to derive the number of jobs in proximity services per inhabitant in French rural municipalities. We first classify the municipalities according to their time distance in minutes by car to the municipality where the inhabitants go the most frequently to get services (called $MFM$). For each set corresponding to a range of time distance to $MFM$, we perform a quantile regression estimating the minimum number of service jobs per inhabitant that we interpret as an estimation of the number of proximity jobs per inhabitant. We observe that the minimum number of service jobs per inhabitant is smaller in small municipalities. Moreover, for municipalities of similar sizes, when the distance to the $MFM$ increases, the number of jobs of proximity services per inhabitant increases.
\end{abstract}

\maketitle

\section*{Introduction}

How many service jobs does each inhabitant of a rural municipality generate in his own municipality? This question is important for the modelling work carried out in the PRIMA European project \cite{PRIMA2011} (The research leading to these results has received funding from the European Commission's 7th Framework Programme FP7/2007-2013 under grant agreement nР212345), dealing with the evolution of rural areas in Europe. In particular, this model aims at incorporating how the growth or decline of municipalities is enhanced by the creation or destruction of these jobs. Indeed, new approaches based on the residential economy point out that the dynamism of rural areas depends significantly on the demand for locally consumed goods and services. We call proximity service these jobs that are generated by the local demand of the municipality, and this paper proposes a method for assessing their number. 

Surprisingly, the literature on the estimation of proximity service job for demographic microsimulation models is very poor. Furthermore the estimation methods proposed are rather crude, for example \cite{Brown2006} proposed a threshold function to create service jobs for one hundred new people. For a direct estimation, the main difficulty is that the available data provide the number of jobs in different categories of services (retail, transportations, various services, public administration, teaching, health and social action) without any information about their relation with the local demand. In the same category, some jobs can depend on the very local market (the municipality), whereas others depend on a wider market of surrounding municipalities or even the whole region. Even the same job of service can be partially devoted to the local customers and partially to a larger market. Therefore, the number of jobs in proximity services can only be estimated indirectly. 

In this paper, we propose to use the minimum requirement approach \cite{Ullman1960} to perform this indirect estimation. This method is usually used for estimating the share of jobs in a given activity \cite{Ullman1960,Brodsky1977}, the employment in touristic activities \cite{Dissart2009,English2000,Leatherman1997} or to compute the regional multipliers giving the propensity to consume locally produced goods \cite{Rutland2007,Woller2002,Persky1994,Moore1975}. In our case, the rationale behind choosing this method is that a large set of municipalities of similar proximity service market always includes some municipalities where the services are only devoted to this local market. These municipalities tend to have the minimum number of service jobs, which gives an estimation of the number of proximity service jobs.

We use two variables to characterise the proximity service market: the municipality size (number of inhabitants) and the offer of services in the neighbourhood. Indeed, the municipality size alone is certainly not sufficient to predict the number of jobs in proximity services because, in our data, the average distance between a municipality and its closest neighbour is about 4 km. Hence there are municipalities that can be very dependent on other ones for their proximity services. We describe the neighbouring offer of services with the time distance by car to the most frequented municipality ($MFM$). The $MFM$ is the municipality where residents from a given municipality usually go to consume services, leisure equipment and facilities that they don't find in their own town. 

In practice, we defined seven municipality sets corresponding to intervals of $tMFM$, the time distance to the $MFM$. In each set, following the minimum requirement approach, we assess the minimum number of jobs per inhabitants with a quantile regression \cite{Koenker1978}, taking as quantile value the first percentile. Indeed, we choose the first percentile (100-quantile) instead of the minimum because the observed data are based on a sample representing a quarter of the population, and the percentile is likely to be more robust to the lack of precision than the minimum. Moreover there is no theoretical justification for using systematically the minimum value \cite{Klosterman1990}. For each of the seven intervals of $tMFM$, we obtain a satisfactory regression predicting the first percentile of service jobs per inhabitant. Moreover, the impact of $tMFM$ corresponds to one's expectations: the municipalities which are close to a $MFM$ have the lowest number of jobs in proximity services per inhabitant and, when $tMFM$ increases, the number of jobs in proximity services per inhabitant increases.

The next section presents the material and methods used for predicting the number of jobs in proximity services per inhabitant. We finally discuss our results.

\section*{Material and methods}

\subsection*{The data from the French statistical office}

This work uses data about municipalities of less than 5000 inhabitants coming from the French Census of 1999, 2006 and 2008 managed by the French Statistical Institute, $INSEE$ and from the French Municipal Inventory of 1999. From this collected data, the Maurice Halbwachs Center or the $INSEE$ makes available to all researchers the following data:

\begin{itemize}
\item The number of inhabitants for each municipality in 1999, 2006 and 2008;
\item The number of jobs in the French tertiary sector (called service jobs) in 1999, 2006 and 2008;
\item The time distance in minutes by car to the most frequented municipality ($tMFM$) in 1999;
\end{itemize}

The $MFM$ is the municipality where residents from a given municipality usually go to consume services, leisure equipment and facilities that they don't find in their own municipality. This variable was obtained in 1999 by asking the following question to the mayor of each municipality \textit{"Where do you go when you need something unavailable in your municipality?"}. The time distance to the $MFM$ is expressed in minutes by car estimated with a average speed/km.  

\begin{figure}
	\begin{center}
		\includegraphics[width=\linewidth]{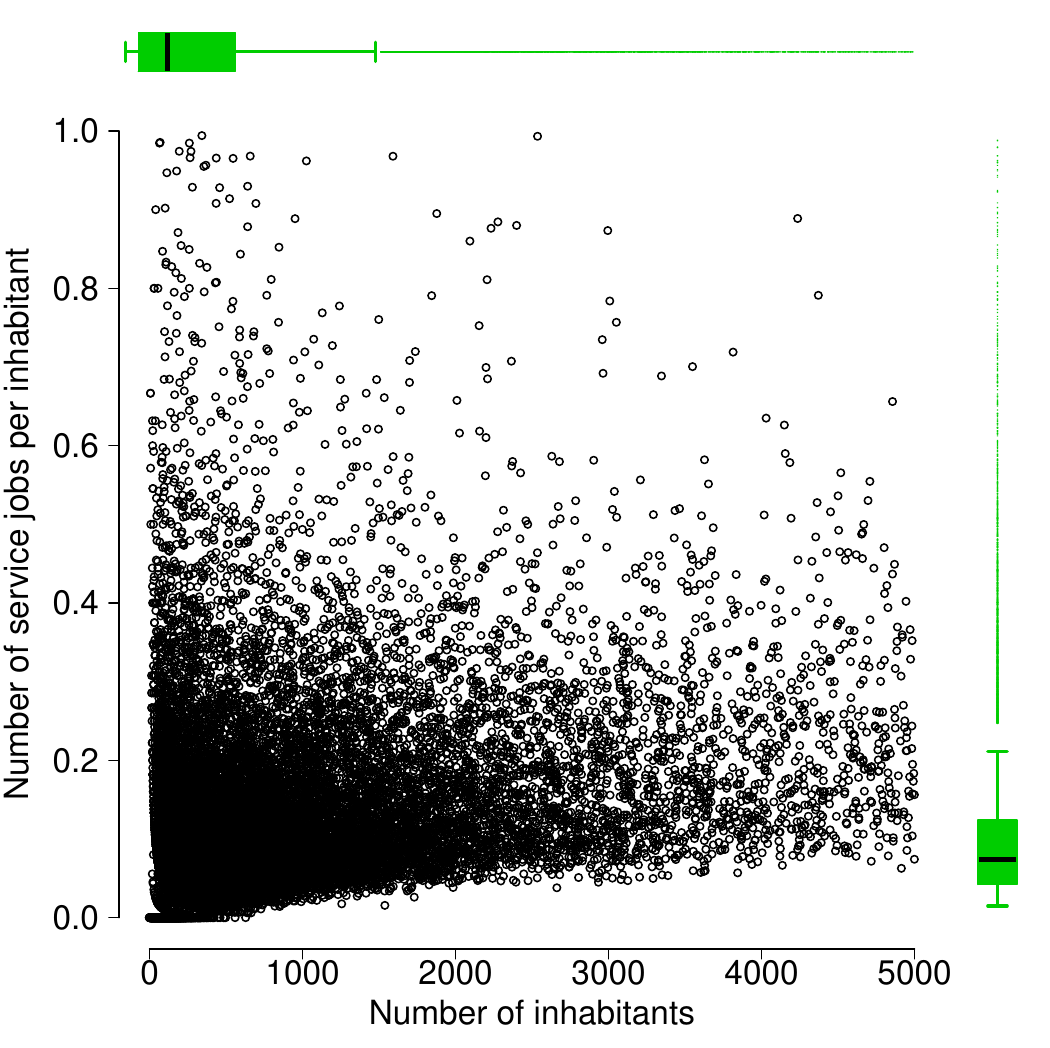}
	\end{center}
	\caption[Number of service jobs per inhabitant]{Number of service jobs per inhabitant function of the number of inhabitants for each municipality in 1999.}
	\label{Figure1}
\end{figure}

\begin{figure}
	\begin{center}
		\includegraphics[width=\linewidth]{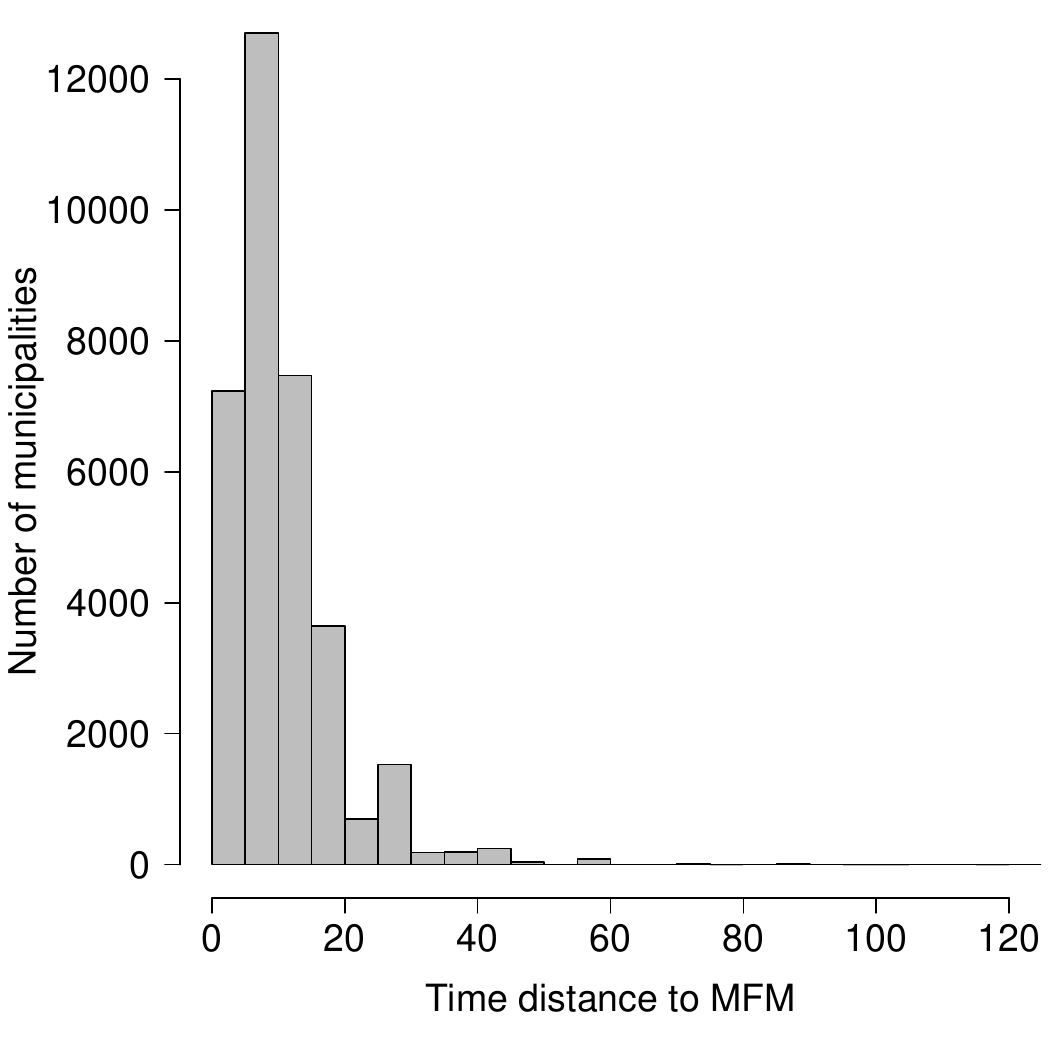}
	\end{center}
	\caption{Histogram of the tMFM in minutes by car in 1999.}
	\label{Figure2}
\end{figure}

\begin{figure*}
	\begin{center}
		\includegraphics[scale=0.6]{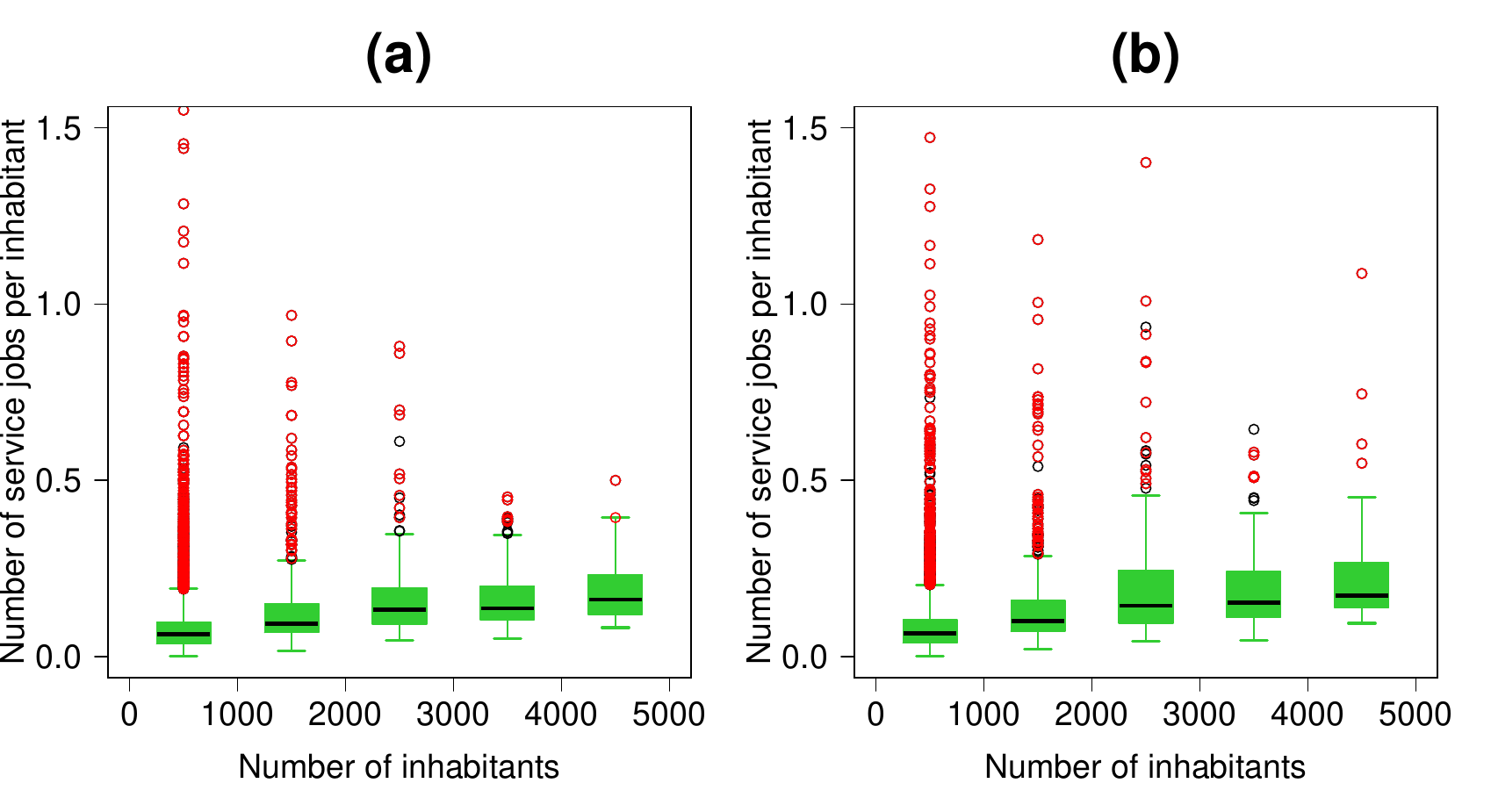}
	\end{center}
	\caption[Box-and-whisker plot of the number of service jobs per inhabitant]{Box-and-whisker plot of the number of service jobs per inhabitant function of the number of inhabitants for tMFM $\in ]0,5]$. The red points represent the common outliers 1999 and 2008. (a) 1999; (b) 2008.}
	\label{Figure5}
\end{figure*}

We observe in Figure \ref{Figure1} that the dataset is mostly composed of small municipalities with a small number of service jobs per inhabitant. We note that the minimum number of service jobs per inhabitant can be expressed by a linear relationship with the logarithm of the number of inhabitants.
We observe in Figure \ref{Figure2} the time distance to the most frequented municipality is mostly between 0 and 20 minutes. The higher is the $tMFM$, the more isolated is the municipality. The $MFM$ of a given municipality is assumed to be the same in 2006 and 2008 as in 1999. In order to check the robustness of this assumption we have highlighted, for a given range of values of $tMFM$, the outliers for the bivariate variable \textit{Number of inhabitants x Number of service jobs} in 1999 and 2008. For different range of values of $tMFM$, the number of outliers is almost the same in 1999 and in 2008, and there is about 80\% of outliers in common between the two time series. We give an example for $tMFM\in]0,5]$ in Figure \ref{Figure5}. Moreover the 20\% "new" outliers in 2008 show a growth of inhabitants that is similar to the one of non-outliers. This does not validate completely the assumption but it reinforces its plausibility.

\subsection*{Model estimate of the number of jobs in proximity services per inhabitant}

In this section, we present the model estimation of the number of jobs in proximity services per inhabitant based on a minimum requirement approach applied to several $tMFM$ intervals.
We assume that the number of jobs in proximity services per inhabitant in a municipality depends not only on the number of inhabitants but also on $tMFM$. Therefore, we define seven sets of municipalities corresponding to intervals of $tMFM$ (values expressed in minutes): $tMFM$$\in]0,5]$, $tMFM$$\in ]5,10]$, $tMFM$$\in ]10,15]$, $tMFM$$\in ]15,20]$, $tMFM$$\in ]20,25]$, $tMFM$$\in ]25,30]$ and $tMFM > 30$. For each of these sets of municipalities we apply a method derived from the minimum requirement approach to estimate the number of jobs in proximity services per inhabitant as a function of the municipality size. 

In general, the minimum requirement approach computes minima on subsets of municipalities of similar sizes, which requires to define these subsets with an appropriate clustering method. We choose to use a quantile regression \cite{Koenker1978}, which does not require to perform this clustering, and yields directly a function estimating the minimum (or a quantile). We choose the first-percentile ($\tau=0.01$) in the regression because our data on the number of service jobs are derived from a sample representing a quarter of the population, and we expect the first percentile to be more robust than the minimum to this lack of precision. 

Let $E$ be the number of service jobs per inhabitant and $P$ the number of inhabitants. We consider the following quantile regression model:
$$E=\beta_0 + \beta_1 \ln{P}+\epsilon$$ where $\beta_0$ and $\beta_1$ are parameters and $\epsilon$ the residual vector. 

With this method, we estimate the number of jobs in proximity services per inhabitant as a function of the municipality size, for each interval of $tMFM$.

\section*{Results}

In this section, we present the results obtained when applying the method on the data from 1999, 2006 and 2008. 

\begin{figure*}
	\begin{center}
		\includegraphics[scale=0.6]{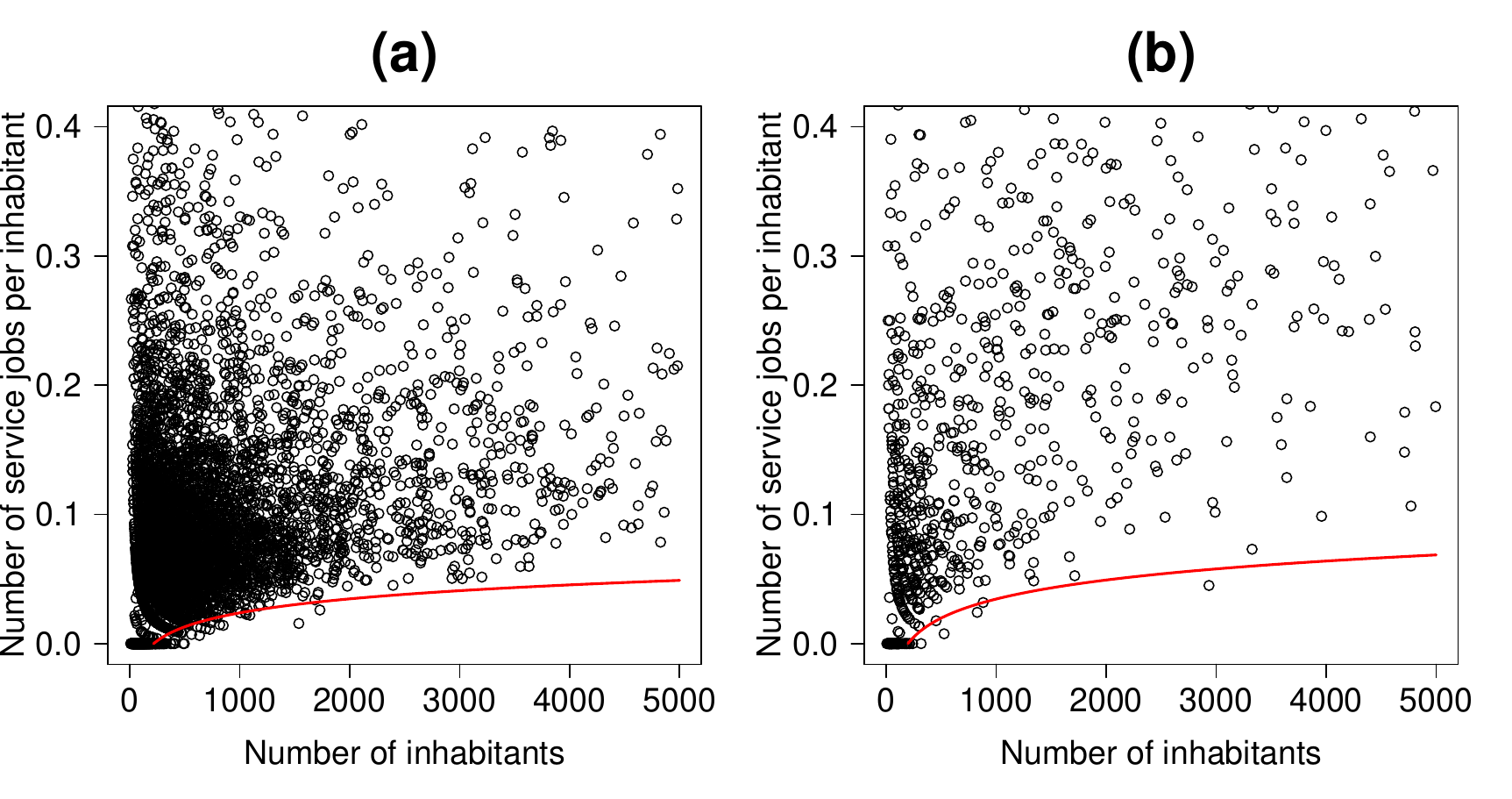}
	\end{center}
	\caption[Number of service jobs per inhabitant for different tMFM]{Number of service jobs per inhabitant function of the number of inhabitants for each municipality in 1999. The line represents the quantile regression line for $\tau=0.01$. (a) tMFM $\in ]0,5]$; (b) tMFM $> 30$.}
	\label{Figure3}
\end{figure*}

The coefficients of the quantile regression for each set of $tMFM$ obtain with the regression quantile $\tau=0.01$ and the 1999 data are presented in Table \ref{Table1}. All the coefficients are significant and the associated standard deviations are quite low. The quantile regression model is significant with one percentile but also with five percentile, ten percentile and the median but we choose the focus on the results for one percentile because we want to be as close as possible to the minimum. Figure \ref{Figure3} shows the relation given by the model for 1999 for $tMFM \in ]0,5]$ and $tMFM > 30$. As we can see on the scatter plots, we obtained a good fit of the model. To assess changes over time in the relationship we have repeated the procedure in 2006 and 2008 (using $tMFM$ from 1999). We note that, for all the $tMFM$ intervals, the slope is positive, and it is the highest for $tMFM > 30$. This implies that the number of proximity service jobs created (or destroyed) is higher in big municipalities than in a small one, when the population evolves, and even higher for municipalities that are far from their $MFM$. 

\begin{table}[!ht]
	\caption{Parameter values and 
	standard deviations (in brackets) of the quantile regression predicting the number of proximity services jobs per inhabitant for the different intervals of tMFM in minutes by 
	car in 1999.}
	\begin{center}
	\begin{tabular}{|l|c|c|}
	  \hline
		\pmb{tMFM} & \textbf{Intercept} & \textbf{Slope}\\
		\hline
		$	]0,5]	$	&	-0.084	(0.0031)	&	0.016	(0.0006)	\\	
    $	]5,10]	$	&	-0.083	(0.0024)	&	0.016	(0.0005)	\\
    $	]10,15]	$	&	-0.079	(0.0014)	&	0.015	(0.0003)		\\	
    $	]15,20]	$	&	-0.094	(0.0025)	&	0.018	(0.0005)\\
    $	]20,25]	$	&	-0.097	(0.0021)	&	0.019	(0.0007)\\
    $	]25,30]	$	&	-0.099	(0.0055)	&	0.019	(0.0012)\\	
    $	>30	$	&	-0.112	(0.0067)	&	0.021	(0.0020)\\	
    \hline
	\end{tabular}
	\end{center}
	\label{Table1}
\end{table}

\begin{figure*}
	\begin{center}
		\includegraphics[scale=0.6]{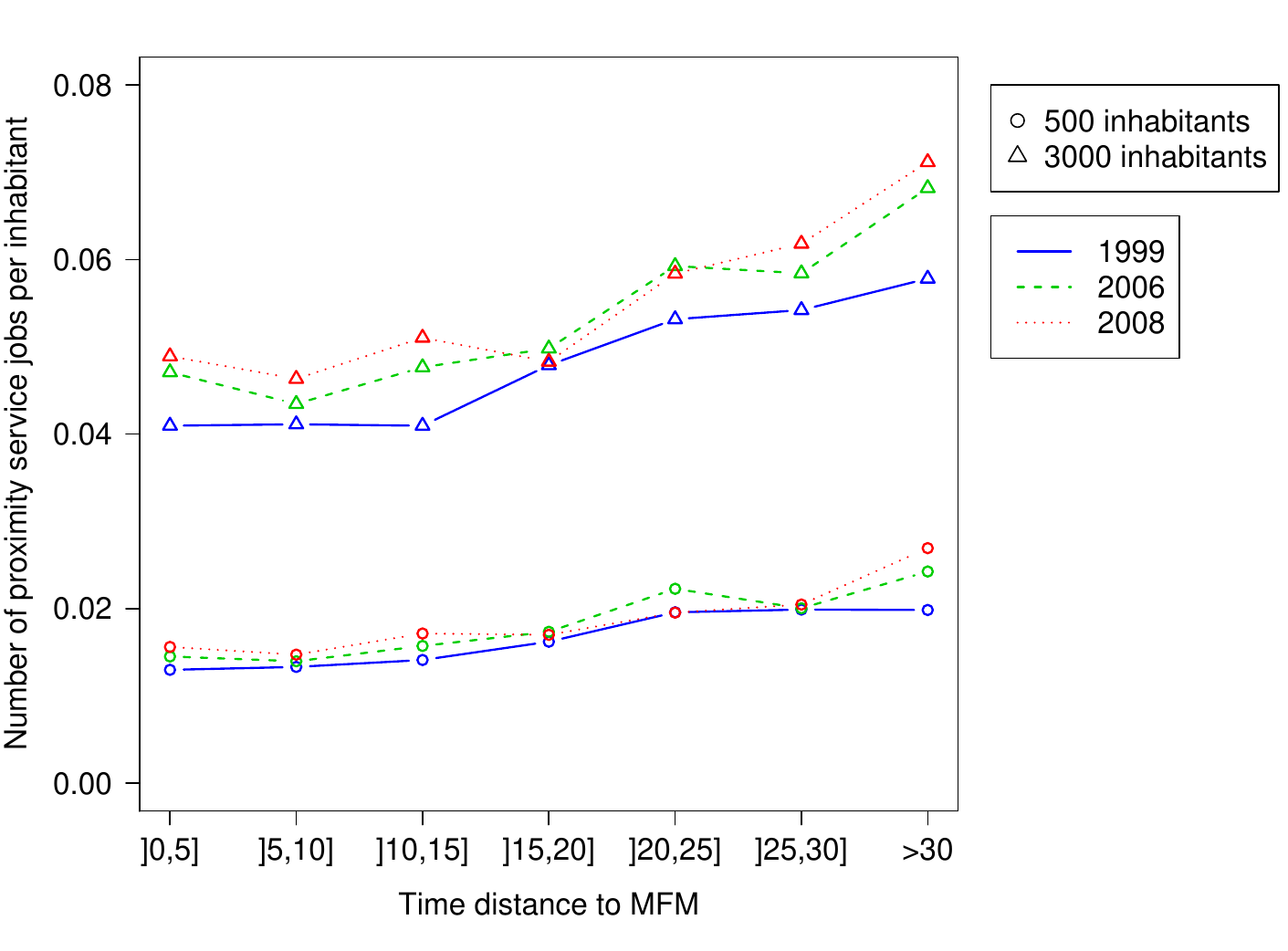}
	\end{center}
	\caption[Number of proximity service jobs per inhabitant function of tMFM]{Number of proximity service jobs per inhabitant function of tMFM interval (min.) (tMFM $\in ]0,5]$, tMFM $\in ]5,10]$, tMFM $\in ]10,15]$, tMFM $\in ]15,20]$, tMFM $\in ]20,25]$, tMFM $\in ]25,30]$ and tMFM $> 30$). Blue solid line for 1999; Green dashed line for 2006; Red dotted line for 2008. Circles: municipality of 500 inhabitants; Triangles: municipality of 3000 inhabitants.}
	\label{Figure4}
\end{figure*}

Figure \ref{Figure4} shows the results for 2006 and 2008 with 1999 for a 500 and a 3000 inhabitants municipality. For each $tMFM$ interval we observe that the number of proximity service jobs per inhabitant tends to increase with time. One can see that the number of proximity service jobs per inhabitant is smaller for $tMFM <15$ and then increases. It is coherent with the results presented in \cite{Mordier2010} which shows the number of service providers is higher in isolated rural area than in suburbs of rural center.The same author shows the number of service providers in rural suburbs is smaller than the on in rural centres (defined as having at least 1500 jobs). The whole form a curve is also coherent with \cite{Hubert2009} who shows that in the rural and weakly urban areas the average daily moving time is 16 minutes in 1994 and 17 minutes in 2008 in France (for those moving by car). 

Finally, within municipalities of 3000 inhabitants, the ones which are $tMFM > 30$ have about 0.02 proximity job services per inhabitant more than municipalities close to $MFM$ ($tMFM < 15$), while this difference is about 0.005 within muncipalities of 500 inhabitants. This suggests that the same population changes in municipalities of 3000 inhabitants, have a significantly higher impact on the proximity service jobs in municipalities far from $MFM$ than in municipalities close to $MFM$. In municipalities of 500 inhabitants $tMFM$ seems to have a weaker impact. 

\section*{Discussion}

We choose the minimum requirement approach for deriving the number of proximity services jobs per inhabitant in French rural municipalities, because it seems reasonable that, in a sufficiently large set of municipalities, some of them have only service jobs for the municipality population itself. Indeed, one can postulate that the long range services are located only in some privileged municipalities. However, we had to adapt the minimum requirement to our problem on three aspects:
\begin{itemize}
	\item Instead of considering the share of jobs in a given activity, we considered the number of jobs per inhabitant. This corresponds better to our assumption that the proximity service jobs depend on the local population.
	\item We performed a series of minimum requirement procedures, corresponding to intervals of time distance to the most frequented municipality. 
	\item Instead of using a discrete model based on a clustering of the municipalities by sizes as in the usual minimum requirement approach, we use a quantile regression \cite{Koenker1978} with as quantile value the first-percentile ($\tau=0.01$).
\end{itemize}

The model yields accurate predictions of the first percentile. It suggests that big municipalities (close to 5000 inhabitants) generate (or destroy) significantly more proximity service jobs than small ones (around 500 inhabitants), for the same growth (or decline) of their population. Moreover, the impact of the time to the most frequented muncipality ($MFM$) corresponds to one's expectations: The municipalities which are close to a $MFM$ have the lowest number of jobs in proximity services per inhabitant, and when the municipality gets farther from the $MFM$, its number of jobs in proximity services per inhabitant increases. Finally, this impact of $tMFM$ on the number of proximity service jobs per inhabitant is significantly higher on big municipalities than on small ones. 

We believe that such results can be interesting for policy makers, who have to make choices for distributing incentives to maintain employment and population in some rural areas. According to our results, the policies will have higher leverage effects in the big municipalities of our sample, especially the one with $tMFM > 30$. Moreover, our results suggest that in municipalities which are close to $MFM$, the population changes are likely to impact also the service jobs in the $MFM$.

\section*{Acknowledgments}

This publication has been funded by the Prototypical policy impacts on multifunctional activities in rural municipalities collaborative project, European Union 7th Framework Programme (ENV 2007-1), contract no. 212345. The work of the first author has been funded by the Auvergne region. We wish to thank Olivier Aznar and Solenn Tanguy for their help and Alexandre Kych from the \textit{Centre Maurice Halbwachs} for his kindness.

\bibliographystyle{unsrt}  
\bibliography{Service}

\end{document}